\documentclass[12pt]{iopart}
\usepackage[dvips]{graphicx,color}

\begin{document}

\title{Asymptotically Exact Localized Expansions for Signals in Time-Frequency Domain}
\author{Aramazd H Muzhikyan$^1$ and Gagik T Avanesyan$^2$}
\address{$^1$ Institute of Radiophysics and Electronics, NAS, Ashtarak, 0203, Armenia\\~~e-mail: avangt@instmath.sci.am}
\address{$^2$ Institute of Mathematics, NAS, Yerevan, 0019, Armenia\\~~e-mail: amuzhikyan@irphe.am}

\begin{abstract}
Based on a unique waveform with strong exponential localization property, an exact mathematical method for solving problems in signal analysis in time-frequency domain is presented. An analogue of the Gabor frame exposes the non-commutative geometry of the time-frequency plane. Signals are visualized using graphical representation constructed.
\end{abstract}

\section{Introduction}

Time-frequency (TF) representation of signals is essential in modern theories. A review of TF analysis methods can be found in  \cite{groc00}, for a brief introduction see \cite{alle04}. Among special mathematical
methods of TF analysis a central role is played by Gabor analysis (see, e.g. \cite{gaba98}).
However, there are difficulties regarding the practical implementations: (i) the existing methods of TF analysis are computationally complex, and (ii) understanding of the TF plane geometry is still poor. It is the aim of present paper to fill this gap.

In TF approach to the sampling, the effects of localization are critical, especially, close to the quantum uncertainty limit. Localization properties of orthonormal bases, when viewed in global context of the whole of the TF plane, are subject to the restrictions imposed by Balian-Low theorem \cite{bene98}. The theorem, as another form of uncertainty relations, reveals deep features of Fourier transformation, in its connection to the noncommutativity of the TF plane.

The fact that this phenomenon has a very general character is known for decades. Parallels with phase-space quantum mechanics and quantum Hall effect describe the phenomenon best (see, e.g. \cite{anto03}). As it is always the case with similar problems in different research areas, they pose a common mathematical problem too. Perhaps the best example in this course is the $kq$-representation constructed by Zak \cite{zak72}, originally for the purposes of the solid-state physics; it was then developed into a general mathematical method used widely in the theory of signals. The first mathematical analysis along these lines goes back to von Neumann \cite{neum55}, within the context of the phase-space quantum mechanics; the idea of elementary signals was put forward by Gabor \cite{gabo46}. Balian \cite{bali81} proved that the bases associated with the Gabor-von Neumann lattice are singular when the orthogonality additional condition is imposed. The similar fact in connection with the quantum Hall effect was established by Low \cite{low85}. A way of circumventing the difficulties in realization of von Neumann's and Gabor's ideas posed by the Balian-Low theorem on localization was shown in \cite{avan04}, by constructing an irreducible representation of the anticommutative subgroup of the Weyl-Heisenberg group, the group of translations-modulations. A unique entire function that naturally arises in that approach leads to exponentially localized basis which solves the problem of local asymptotic approximations to entire functions, in Hilbert space metrics \cite{avan08}. Due to its unitary invariant form, the approach can be applied directly to the signal analysis in TF plane.

Conventionally, sampling of the signal represented by the function $f(t)$ means an approximation to the function using a set of its values $f_k=f(t_k)$ at the given set of the instants $t_k$. On the other hand, there is no reasonable way to deal with the integrals $f(t_k)=\int \delta(t-t_k)f(t)dt$ using real sampler. We introduce a different method of discretization, based on a function exponentially localized both in the time and the frequency, instead of using Dirac delta function. The method is a realization of the representation that uses Gabor's idea of `elementary' signals, which helps to overcome the difficulties with classical Gabor decomposition, connected with the incompatibility of orthogonality and localization, simultaneously, according to the Balian-Low theorem. Close to our approach is the one using Wilson bases \cite{wils00}.

 An introduction to the method of asymptotically exact expansions in TF plane with strong exponential localization property is provided below, followed by its applications to signal analysis, in a form of several basic examples, and using graphical representation constructed.

\section{Mathematical Preliminaries: the Unique Waveform and Asymptotically Exact Expansions}
\label{sec:mathematical}
An insight into the noncommutative geometry of the TF plane, as viewed from the position of signal processing, may be found by taking into account the fact that translations and modulations form a non-commutative continuous group: the Weyl-Heisenberg group. Its maximal commutative subgroup `forming' the von Neumann lattice, called also Gabor frame, is subject to the restrictions imposed by the Balian-Low theorem. Consideration of the anticommutative subgroup of the Weyl-Heisenberg group, with double-density lattice, shows a way to deal with the singularities present. The subgroup under consideration is generated by translations, $T_a: f(t)\rightarrow f(t+a)$, and modulations, $T_b: f(t)\rightarrow \exp(ibt)f(t)$, with $a=b=\sqrt{\pi}$, and anticommutation relation
\begin{equation}
T_aT_b+T_bT_a=0
\label{eq:anc}
\end{equation}
holds. The details of this approach are given in \cite{avan04}. Because of the incompatibility of localization and orthogonality, one has either to use the description in terms of nonorthogonal subspaces, or to tolerate very weak localization (as for the `$sinc$' function), and infinite uncertainty products $\Delta \omega \Delta t$.

The overall complication thus reduces to the consideration of four mutually nonorthogonal subspaces of the Hilbert space, represented by the respective four sublattices in the TF plane. Within each of the four subspaces there exists an orthonormal basis with strong exponential localization property, represented by the shifts of a single waveform,
\begin{equation}
a(t)\propto\frac{1+2\sum\limits_{n=1}^{\infty}(-1)^n\alpha_ne^{-\pi n-2\pi n^2}\cosh(2nt\sqrt{\pi})}{e^{\frac{t^2}{2}}\sqrt{\vartheta_3(t\sqrt{\pi})}}.
\label{eq:ati}
\end{equation}
The waveform is basic to our construction. The factor $e^{-\pi n}$ in the sum above is responsible for strong exponential localization. The function (\ref{eq:ati}) coincides with its own Fourier transform. (The same property has Hermite functions under proper scale, and that symmetry is exploited in our approach, see the next section.) The four sublattices correspond to the subspaces $M_\mu \subset H$, $\mu=1,2,3,4$, which i) are invariant with respect to $T_a^2$ and $T_b^2$, ii) $M_3$ has an orthonormal basis generated from the base waveform by the all combinations of the shifts $T_a^{2m}$ and $T_b^{2n}$ with integer $m$ and $n$ (the third sublattice), and iii) $M_4=T_aM_3$, $M_2=T_bM_3$, and $M_1=T_aM_2=T_bM_4$.

The function $\vartheta_3(x)$ that enters the definition above corresponds to the so called lemniscate case of the theory of elliptic functions, and by $\vartheta_3(x)$ we mean Jacobi theta function $\vartheta_3(x,e^{-\pi})$; explicitly,
\begin{equation}
\vartheta_3(x)=1+2\sum\limits_{n=1}^{\infty}e^{-\pi n^2}\cos(2nx).
\label{eq:th3}
\end{equation}
The coefficients $\alpha_n$ are defined as coefficients of the Fourier series expansion for the inverse square root of Jacobi theta function $\vartheta_4(x)=\vartheta_3(x+{\pi}/2)$:
\begin{eqnarray}
\frac{1}{\sqrt{\vartheta_4(z)}}\propto1+2\sum\limits_{n=1}^{\infty}\alpha_ne^{-\pi n}\cos(2nz).
\label{eq:theta}
\end{eqnarray}
The first six values of $\alpha_n$ are: \{0.501 052, 0.375 586, 0.312 961, 0.273 833, 0.246 446, 0.225 907\}. It may be said that the numbers $\alpha_k$ are structure constants of the TF plane with respect to the symmetry group of that plane, the Weyl-Heisenberg group, that are generated by its anticommutative subgroup.

\begin{figure}[!t]
\centering
\scalebox{0.4}{\includegraphics{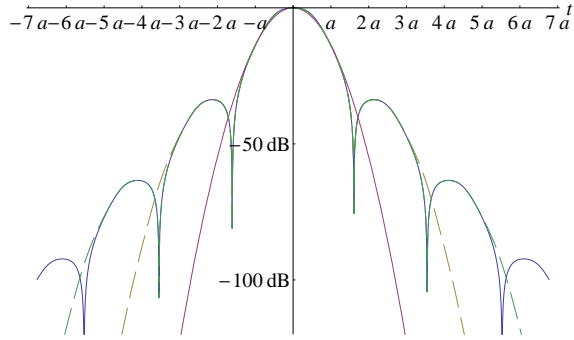}}
\caption{Power attenuation-both with the time and the frequency, for the signal (\ref{eq:at}), as compared to that for the Gaussian signal (the parabola). Also shown are the three- and the five-term approximations to the signal (dashed lines).}
\label{fig:at}
\end{figure}

For practical calculations, the function (\ref{eq:ati}) can be rewritten in a form
\begin{equation}
    a(t)\propto\frac{e^{-\frac{t^2}{2}}+\sum\limits_{n=1}^{\infty}(-1)^n\alpha_ne^{-\pi n}(e^{-\frac{(t-2n\sqrt{\pi})^2}{2}}+e^{-\frac{(t+2n\sqrt{\pi})^2}{2}})}{\sqrt{\vartheta_3(t\sqrt{\pi})}}.
\label{eq:at}
\end{equation}
Comparison of the function (\ref{eq:at}) with the Gaussian waveform is shown in Fig.\ref{fig:at}. Also shown are the three- and the five-term approximations to the function (\ref{eq:at}), in order to demonstrate the superconvergence of the sum in the nominator.

The phases of the basis vectors are fixed as follows:
\begin{equation}
a_{m,n}(t)=e^{inb(t+ma/2)}a(t+ma),\ \ m,n = 0, \pm 1, ...\ .
\label{eq:amnt}
\end{equation}

The most important feature clearly seen in the Fig.\ref{fig:at} is the strong exponential decay. The fact of the linear relation between the time and the frequency scales with attenuation is basic to our asymptotically exact reconstruction method, which follows
from the resolution of identity for the Hilbert space of signals with finite energy. The respective scalar products are calculated as integrals

\begin{equation}
f_{m,n}=\int f(t)\overline{a_{m,n}(t)}dt,\ \ m,n = 0, \pm 1, ...\ .
\label{eq:fmn}
\end{equation}

The same fundamental fact of strong exponential localization allows to use finite sums as asymptotically exact approximations,

\begin{equation}
f(t)\approx\frac{1}{2}\sum {f_{m,n}a_{m,n}(t)},
\label{eq:reconstruct}
\end{equation}
within the rectangle ${|m|} \leq M, {|n|} \leq N$. The finite sets $\{f_{m,n}\}$ would represent the asymptotic expansion, understood in the sense of Hilbert space metrics.

Because of nonorthogonality between the four sublattices, the energy of the signal is counted effectively twice so that
\begin{equation}
\frac{1}{2}\sum|f_{m,n}|^2=\int| f(t)|^2 dt,\ .
\label{eq:norm}
\end{equation}

With the equations (\ref{eq:fmn}) and (\ref{eq:reconstruct}), it doesn't matter which particular representation of the signal under consideration is used, either the time or the frequency one. The numbers $M$ and $N$ are chosen such that the energy of the signal is well approximated according to the (\ref{eq:norm}) with respect to the noise level. Further details of the basis construction, and its properties, can be found in the two papers \cite{avan04} and \cite{avan08}.

Hilbert space vectors corresponding to the signals, in the basis under discussion are conveniently represented by complex matrices indexed by discrete time and frequency $m$ and $n$, so that the corresponding "matrices" themselves play the role of the coordinates of the signal in the functional space. The "matrices" based on expansion formulae (\ref{eq:fmn}) and (\ref{eq:reconstruct}) can be represented in a graphical form, by using polar decomposition of complex numbers--the elements of the matrix, with semicircles positioned in the proper cells in TF plane, as described below.

\section{Graphical Representation: Sampling and Reconstruction of Commonly Used Signals}
\label{sec:graphical}

\begin{figure}[!t]
\centering
\scalebox{0.4}{\includegraphics{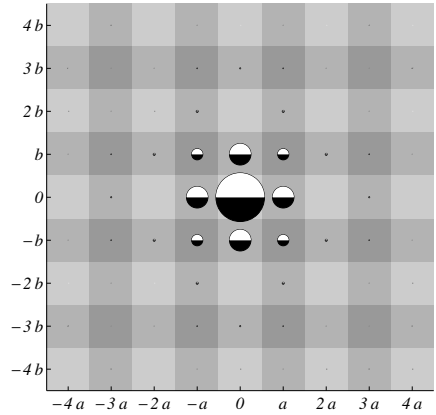}}
\caption{Graphical representation for the function $a(t)$ according to (\ref{eq:fmn}) and (\ref{eq:reconstruct}). Time and frequency axes are horizontal and vertical, respectively. }
\label{fig:atomic}
\end{figure}

Graphical representation for the function (\ref{eq:at}) itself that is based on asymptotic expansion of the type (\ref{eq:reconstruct}) is shown in Fig.\ref{fig:atomic}. This example is basic: it makes explicit the noncommutative structure of the plane that is reflected in the nonorthogonality between the four subspaces of the Hilbert space of signals, which are represented by the respective four sublattices marked by alternating levels of grayscale. Recall that the symmetry of the four sublattices is described by the anticommutative pair of $T_a$ and $T_b$ (\ref{eq:anc}). The expansion is characterized by the complex numbers--the amplitudes (\ref{eq:fmn}), that are attributed to each site of this double overcomplete union of sublattices.
The interpretation of the graphs introduced is straightforward: the sites of the four sublattices represent bases, orthonormal within each of the four subspaces. The sites with coordinates $\{ma,nb\}$ are attributed to the functions $a_{m,n}(t)$ (\ref{eq:amnt}). The radii of the circles are equal to the absolute values $|f_{m,n}|$, and rotations represent their phases, $Arg(f_{m,n})$; black semicircle down means zero phase. For instance, in Fig.\ref{fig:atomic} the amplitude at the origin, $f_{0,0}$, is real and positive. The nonzero amplitudes on the adjacent lattices reflect the nonorthogonality of the four subspaces of the Hilbert space of signals, which is inevitable in the view of the noncommutativity of the TF plane. Considered here are expansions for normalized signals (signals of unit energy); the portion of the energy of the signal at the given site is proportional to the respective circle area. Symmetry  with respect to Fourier transformation is reflected in the symmetry with respect to the time and the frequency axes. Graphical representation described here demonstrates the functionality of our method based on expansion formulae (\ref{eq:fmn}) and (\ref{eq:reconstruct}).
\begin{figure}[!t]
\centering
\scalebox{0.4}{\includegraphics{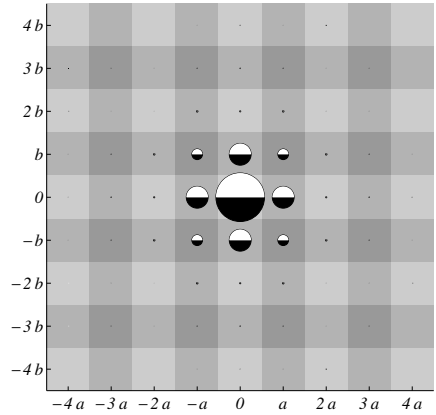}}
\caption{Gaussian waveform represented by its asymptotic expansion in the TF plane.}
\label{fig:gaussian}
\end{figure}

\begin{figure}[!t]
\centering
\scalebox{0.4}{\includegraphics{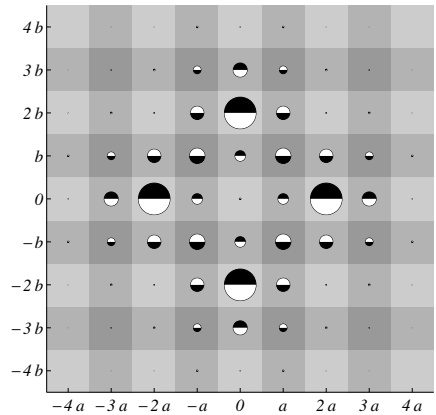}}
\caption{The difference between the base waveform (\ref{eq:at}) and the Gaussian waveform, both normalized. The radii of the circles are shown with magnification 30, to expose minor differences that cannot be seen clearly in Fig.2 and Fig.3.}
\label{fig:compare}
\end{figure}

A convenient tool for the representation of common waveforms are the expansions in Hermite functions; the latter form orthonormal basis in the space of signals with finite energy. On the other hand, it is advantageous to use their asymptotic expansions of the type (\ref{eq:reconstruct}) for the representations of the waveforms translated-modulated. The interrelation between the two bases, the one, represented by the four sublattices, and the other, represented by the set of Hermite functions, once established, offers a convenient tool for signal representation and analysis: the advantages of the latter with respect to the signals 'centered' about the origin now can be combined with the translation invariance of the former, both with the time and the frequency.

\begin{figure}[!t]
\centering
\scalebox{0.4}{\includegraphics{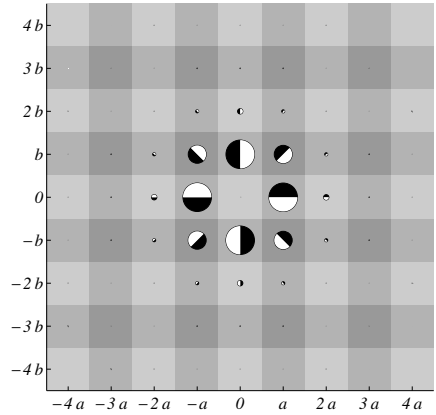}}
\caption{Gaussian monocycle waveform in the TF plane.}
\label{fig:monocycle}
\end{figure}

\begin{figure}[!t]
\centering
\scalebox{0.4}{\includegraphics{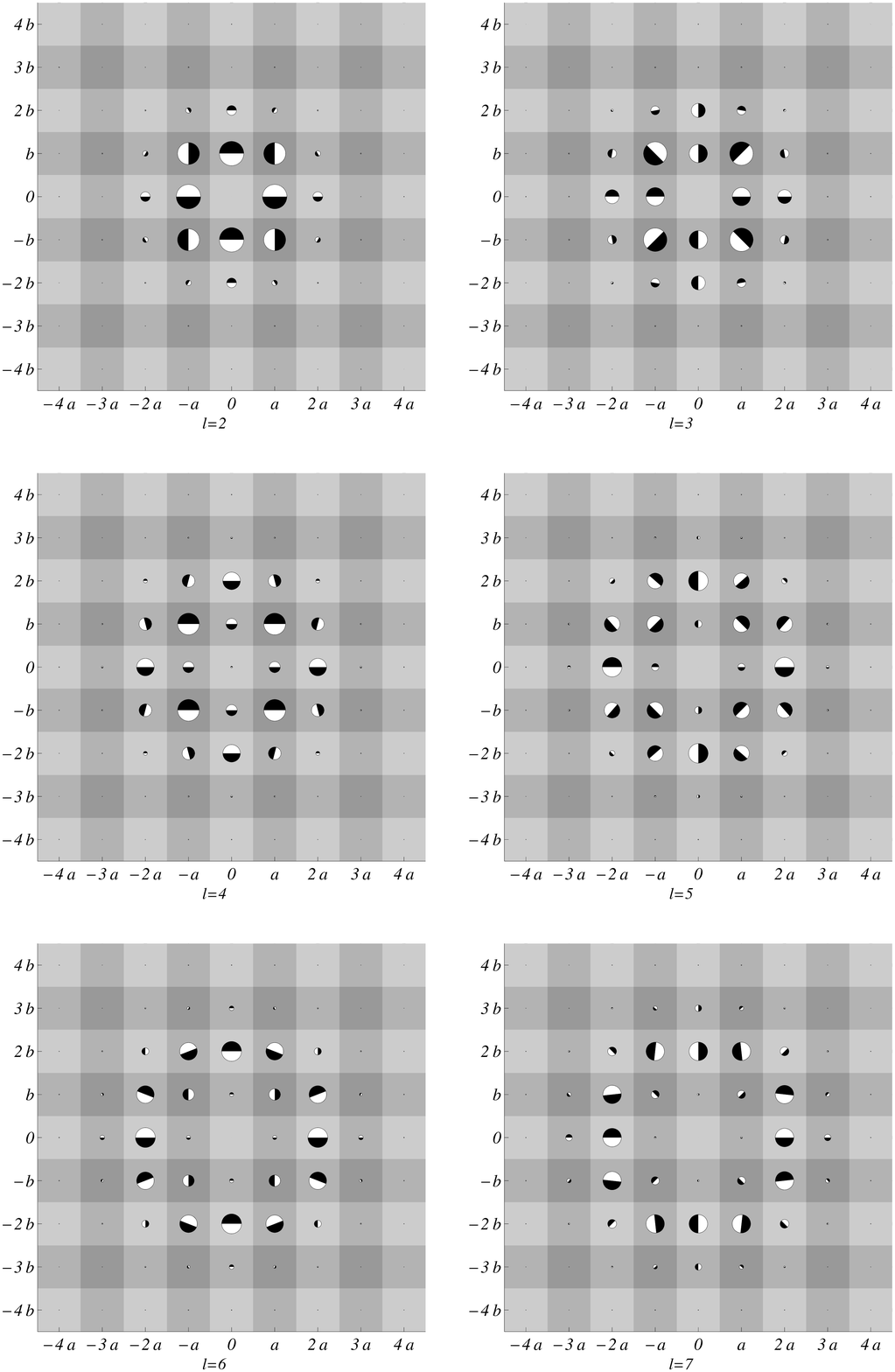}}
\caption{Hermite functions of higher orders in TF plane.}
\label{fig:hermite}
\end{figure}

Signals most localized in the TF plane are represented by the Gaussian waveform: its uniqueness is guaranteed by the minimization of the Heisenberg uncertainty relation. Less localized are the so-called Gaussian monocycles, and the pulses of higher order, represented in terms of quantum mechanics as higher excited states of a harmonic oscillator, and their hierarchy constitutes a good classification scheme for the degree of localization. The character of localization of the Gaussian signal is seen clearly in the Fig.\ref{fig:gaussian}. The main part of the energy of the signal is concentrated on the central sites of the lattice, and the other amplitudes nearly vanish. The rotational symmetry of the expansion is an effect of the symmetry with respect to the Fourier transformation.  The position and labeling of the sublattices coincide with those in the Fig.\ref{fig:atomic}. The minor differences between $a(t)$ and the Gaussian waveform that reflect the orthonormality of the former on its own sublattice, and the absence of such for the latter, are plotted in Fig.\ref{fig:compare}. Notice that the difference maxima are on the sites $(\pm 2a,0)$, and $(0,\pm 2b)$, which belong to the `central' sublattice, and that the central difference is relatively very small.
The next waveform considered is the so called Gaussian monocycle, Fig.\ref{fig:monocycle}. It is less localized than the Gaussian one, as  expected. The symmetry of the figure is due to the symmetry of the waveform with respect to the Fourier transformation, as for the Gaussian waveform. In contrast with the Gaussian one, the central amplitude is exactly zero, and the amplitudes on the frequency axis are purely imaginary; the phases of the amplitudes on the other sites vary in a symmetric manner.
The two waveforms considered above are actually the zeroth and the first in the set of Hermite functions, which constitute a basis in the Hilbert space of signals, as it was mentioned above. Since the interrelation of the two complete sets is of primary interest, the other Hermite functions will be considered too. The hierarchy of the eigenvectors of quantum harmonic oscillator Hamiltonian can represent the degree of localization, parameterized by the order of the polynomial, ${l}$. Normalized Hermite functions are

\begin{figure}[!t]
\centering
\scalebox{0.6}{\includegraphics{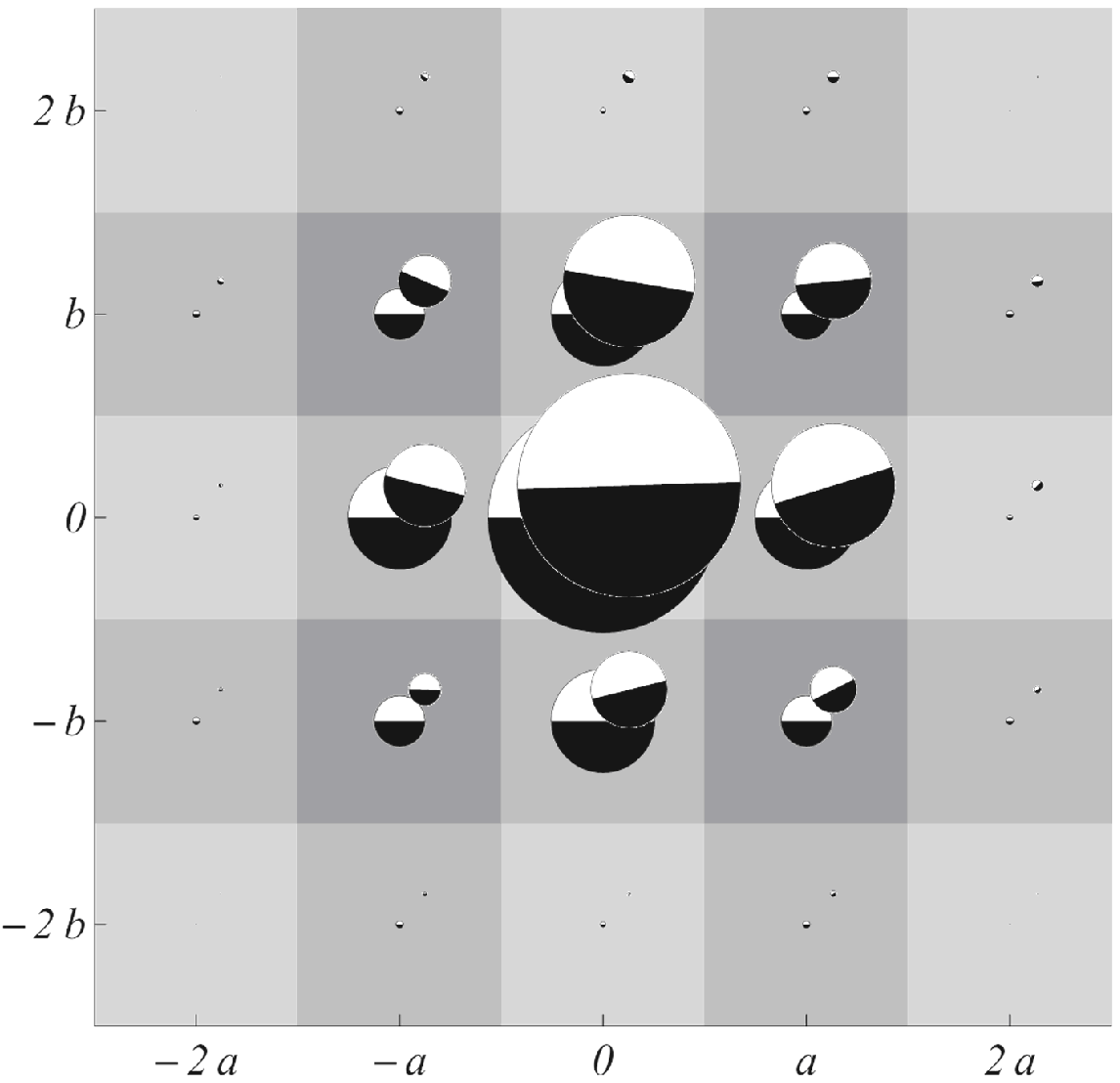}}
\caption{Standard signal at a `random' position in TF plane so close to the origin that the uncertainty product is much less than the uncertainty limit imposed by the Heisenberg relation. The signal expansion is shifted intentionally according to the values of the $\Delta t$ and $\Delta \omega $.}
\label{fig:shift}
\end{figure}

\begin{equation}
{u_l}(t)=\frac{1}{\sqrt{l!2^l\sqrt{\pi}}}e^{-t^2/2}H_l(t),\ \ l = 0, 1, 2, ...
\label{eq:hermitef}
\end{equation}
and $H_l(t)$ are Hermite polynomials
\begin{equation}
H_l(t)=(-1)^le^{t^2}\frac{d^l}{dt^l}e^{-t^2}.
\label{eq:hermitep}
\end{equation}

The main features of the graphical representation for the two waveforms considered above, their symmetry and localization, are persistent with the whole hierarchy. Hermite functions of higher orders are shown in Fig.\ref{fig:hermite}. By establishing the interrelation between the two bases, the one, localized about the origin, and the other, uniformly spaced in the plane, we show the way to localized asymptotic approximations applicable to any signal.

With the property of the strong exponential localization of the waveform (\ref{eq:at}), the expansion method presented is equally applicable to arbitrary displacements in the TF plane. In many signal processing applications the recognition of a standard signal displaced in the TF plane, i.e. translated and modulated, is sought. Calling it $a(t)$, the problem reduces to calculations with single function of the displacement, $C(\xi,\eta)$, that is represented by the integral
\begin{equation}
C_{m,n}(\xi,\eta)=\int e^{i\eta (t+\xi/2)}a(t+\xi)\overline{a_{m,n}(t)}dt.
\label{eq:radar}
\end{equation}
It is clear that the restrictions of the uncertainty relations would not affect the recognition, within our method using expansions of the type (\ref{eq:reconstruct}). On the other hand, the set of scalar products represented by the integral (\ref{eq:radar}) reveals noncommutative geometry of the TF plane in a manner different from that of the Fig.\ref{fig:atomic}. Fig.\ref{fig:shift} presents an example: the standard signal at `random' position in the TF plane very close to the origin, which gives another insight into the structure of the space of signals in quantum uncertainty limit.

\section{Discussion and Conclusions}
\label{sec:remarks}

The sample of the signal represented by the function $f(t)$, when understood directly, is the value of the function at the given instant $t_0$.  Clearly, such understanding is not an `operational' one. By assuming the form of integrals (\ref{eq:fmn}) for the samples, the concept of sampling in TF plane looks much more transparent; and it is operational. In this course, the conceptual aspect may be formulated simply as a rule to deal with operational quantities. Furthermore, such an approach is a realization of the representation that uses Gabor's `elementary' signals, which helps to overcome the difficulties with classical Gabor decomposition, connected with the incompatibility of orthogonality and localization, simultaneously, according to the Balian-Low theorem. The inevitable complications caused by the partial nonorthogonality then reflect the noncommutativity features of the TF plane. It should be stressed that the scalar products represented by (\ref{eq:fmn}) have invariant meaning.

In conclusion, (i) an asymptotically exact mathematical method for localized expansions of signals in TF domain is presented; (ii) based on the method, a graphical representation for signals in the TF plane is introduced; (iii) noncommutative geometry of the TF plane is visualized; (iv) the interrelation with the Hermitian basis is established, in terms of the graphical representation constructed.

\section*{Acknowledgments}
The authors would like to thank N. Engibaryan, A. Hakhoumian, L. Murza and T. Zakaryan for valuable discussions.

\end{document}